\documentstyle[aps,pra,epsfig]{revtex}
\newcommand{\W}{\Omega}

\newcommand{\bea}{\begin{eqnarray}}
\newcommand{\eea}{\end{eqnarray}}

\newcommand{\ds}{\displaystyle}

\newcommand{\be}{\begin{equation}}
\newcommand{\ee}{\end{equation}}

\begin{document}

\title{Broadband optical gain via
interference in the free electron laser: principles and
proposed realizations}

\author{Yuri V. Rostovtsev$^{1,2}$, 
Gershon Kurizki$^{3}$, and 
Marlan O. Scully$^{1,2}$}

\address{$^{1}$ Department of Physics, Texas A\&M University,\\
College Station, Texas 77843-4242 \\
\vspace*{2mm}$^{2}$ Max-Planck-Institut f\"{u}r Quantenoptik,\\
Hans-Kopfermann-Str. 1, D-85748, Garching, Germany,\\
\vspace*{2mm}$^{3}$ Chemical Physics Department, \\
Weizmann Institute of Science,\\
Rehovot 76100, Israel}
\date{\today}
\maketitle

\begin{abstract}
We propose experimentally simplified schemes of an 
optically dispersive interface region between two coupled 
free electron lasers (FELs), 
aimed at achieving a much broader gain bandwidth than in a
conventional FEL or a conventional optical klystron composed of
two separated FELs. The proposed schemes can {\it universally} 
enhance the gain of FELs, regardless of their design when operated 
in the short pulsed regime.
\end{abstract}

%\narrowtext
%\twocolumn

\section{Introduction}

Free-electron lasers (FELs) convert part of the kinetic energy 
of nearly-free relativistic electrons into coherent radiation 
%\cite{FEL,madey78,Madeythe}.
\cite{madey78,FEL,Nikon,Madeythe}.
They can be of two types:
1) devices wherein the electrons are accelerated in a spatially 
periodic magneto-static, electrostatic,  or electromagnetic field, 
called a wiggler or an undulator; 
2) devices wherein the electrons are unperturbed, but instead 
the laser wave is subject to dispersion, as in Cherenkov transition 
radiation.
 
In the classical description of FELs based on wigglers, 
the combined effect of the wiggler and laser fields 
yields a pondermotive potential which causes bunching of 
the electrons along the wiggler axis. This bunching is 
associated with the gain or loss of energy by electrons, or, 
equivalently, their axial acceleration or deceleration, 
depending on the phase between their transverse motion and 
the laser wave. The oscillation 
of electrons in the pondermotive potential is described by 
the pendulum equation, which can yield near-resonant gain, 
provided the electron velocity allows it to be nearly-synchronous 
i.e., maintain a slowly-changing phase, with this potential. 
The near-synchronism condition is for the electron velocity 
to be near the resonant velocity
\be
v_r = {\nu\over k_L + k_W},
\ee
where $\nu$ is the laser frequency, and $k_{L(W)}$ are 
the laser (wiggler) wavevectors. 
The electrons whose velocities are above $v_r$ contribute 
on average to the small-signal gain (radiation emission), 
and those whose velocities are below $v_r$ contribute 
on average to the corresponding loss (radiation absorption). 

This results in an {\em antisymmetric} dependence of the small-signal 
standard gain $G_{st}$ on the deviation of the electron velocity 
$v$ from the resonant velocity $v_r$. Such dependence has been 
thought to be a fundamental consequence of 
the Madey gain-spread theorem \cite{madey78,FEL,Nikon}, 
which states that the gain lineshape is antisymmetric, since it is 
proportional to the derivative of the symmetric spontaneous-emission 
lineshape, which is a sinc$^2$ function of $(v - v_r)$. 
This gain lineshape allows for net gain only if the initial 
velocity distribution is centered above $v_r$, 
which is often called {\em momentum population inversion}. 
In other words, this gain lineshape 
restricts the momentum spread 
to values comparable with the width of the positive (gain) part of $G_{st}$ in
order to achieve net gain.
This width severely limits 
the FEL gain performance at short wavelengths \cite{FEL,Nikon,Madeythe}.  

In a variant of the FEL, composed of two wigglers separated 
by a drift region between them (see Fig.~1) 
firstly suggested by Vinokurov \cite{vinokurov77}, 
known as an "optical klystron" \cite{ok}, 
the first wiggler serves to "bunch" the electron phases, 
which then acquire favorable values in the drift region 
between the wigglers, and finally yield enhanced gain 
in the second wiggler. However, in this case the width of the gain 
region is proportionally narrower, and this makes 
the restrictions on the velocity spread even more severe. 

In an attempt to overcome the adverse effect of momentum 
spread on FEL gain, we have put forward ideas inspired 
by lasing without inversion (LWI) \cite{kocharovskaya92} in atomic systems, 
namely, the cancellation of absorption by interference 
in the gain medium. The analogous schemes proposed by us for FELs 
\cite{kurizki93prl,nikonov96oc,sherman95prl,nikonov96pre} 
involve two wigglers coupled by a specially designed drift region, 
which yields a gain curve that differs substantially from that of 
an optical klystron. 
In an optical klystron, 
the electrons drift between the wigglers in a dispersive region and, thereby, 
acquire a phase shift, relative to the pondermotive potential,  
that  
{\it increases} with the deviation of the electron velocity 
from $v_r$. 
This is in contrast to our schemes 
\cite{nikonov96pre}, in which electrons in the drift region are 
magnetically deflected so as to acquire a dispersion that 
is just opposite to the 
one for optical klystron: The introduced phase shift 
$-({\mathbf k}+{\mathbf k_W})({\mathbf v}+{\mathbf v_r})T$ 
($T$ being the mean interaction time in the wiggler)
{\it decreases} 
with increasing deviation of the electron velocity 
from $v_r$.
This cancels the interaction phase picked up in the first wiggler, 
i.e., bunching is reversed. 
In addition, electrons with {\em initial} velocities 
$v < v_r$, 
which contribute on average to loss (absorption),  
are given a phase shift of $\pi$, in order to cause
destructive interference   
with electrons that contribute to loss in the second wiggler. 
In the resulting gain curve, the usual absorptive part (below resonance) 
is eliminated, whereas the gain part above resonance  
is doubled \cite{nikonov96pre}. 
This implies that net gain is obtained in such schemes even 
from beams with a very broad (inhomogeneous) momentum spread, 
whence we named them FEL Without Inversion (FELWI), analogously to atomic LWI. 

The previously proposed FELWI schemes 
\cite{kurizki93prl,nikonov96oc,sherman95prl,nikonov96pre} 
may open new perspectives for short-wavelength FELs,  
provided that the technical challenges associated with
magnetic field designs for the drift region are adequately met. 
In this paper we propose a considerably simpler variant of 
such schemes, aimed at extending the {\em optical gain 
bandwidth\/} in FELs. The proposed setup (Fig. 1) is equivalent to 
the previously proposed extension of momentum spread capable 
of gain, by phase shift manipulations, yet it involves only optical 
(laser) phase shifts in the drift region (Sec.~II). 
These phase shifts are much easier to
manipulate, since they require only {\em linear\/} optical 
elements - prisms or Bragg mirrors, etc. (Sec.~III). 
We conclude that the proposed scheme is {\it universal}, i.e., applicable
to FELs regardless of their wiggler design. It
may substantially enhance the 
FEL performance in the pulsed regime (Sec.~IV). 

\section{Principles of the broadband gain mechanism in FELs}
\subsection{General formula for small-signal gain in two 
interfering wigglers}

The dynamics of an electron interacting with 
the laser field in wiggler I or II,  
is expressed by the pendulum equations  
\cite{FEL}   
%%%%%%%%%%%%%%%%%%
%\begin{eqnarray}
%\displaystyle \frac{d\psi}{dt} & = & \Omega \\[2mm] 
%\displaystyle \frac{d \Omega}{dt} & = & a \sin\psi,
%\end{eqnarray}
\begin{equation}
\displaystyle \frac{d\psi}{dt}  =  \Omega, \;\;\; 
\displaystyle \frac{d \Omega}{dt}  =   a \sin\psi,
\label{pendulum}
\end{equation}
%%%%%%%%%%%%%%%%%%%%%%%%%%%%%
%\begin{equation}
%%\displaystyle\left\{ 
%\begin{array}{rcl}
%\displaystyle \frac{d\psi}{dt} & = & \Omega \\[2mm] 
%\displaystyle \frac{d \Omega}{dt} & = & a \sin\psi,
%\end{array}
%%\right.  
%\label{pendulum}
%\end{equation}
%
where 
\be
\W = q_z(v_z - v_r), \;\; 
\psi = - \Delta\nu t + q_z z + q_x x + \phi, \;\;
q_x = k_L \sin\theta, \;\; 
q_z = k_L \cos\theta + k_W.  
\ee
The coupling constant 
\begin{equation}
a= \left( k_x^2 + \frac{q_z^2}{\gamma_r^2} \right) 
\frac{2 e^2 A_W A_L}{p_z^2}.
\end{equation}
is proportional to the laser field amplitude and will be used as the
perturbation parameter in the small signal regime. 

The dynamics and resulting gain depends on the detuning $\W$,
which is a function of the laser frequency and the initial velocity 
$v_z$. A small change in the laser frequency by $\delta\omega$ 
or in the electron velocity by $\delta v_z$ produces a {\it similar 
effect} on $\W$, and 
thus on the FEL dynamics and gain (Fig.~\ref{fel-dw}a,~b). 
Here we assume that the electron beam has a narrow 
distribution of electron velocities $v_z$, 
and are mainly interested in the gain dependence on the
laser frequency variation $\delta\omega$. 
The detuning then depends on laser frequency as 
\begin{equation}
\Omega =q_zv_z-\delta \omega \left( 1-{\frac{v_z}c}\right).
\end{equation}

Equations (2)-(4) are the basis of our consideration of uniform wigglers. 
In the ultrarelativistic limit, small changes of the energy, momentum, 
detuning, and velocity are proportional to each other, so that in order 
to calculate the gain we need only calculate the change 
in the detuning of the electrons upon averaging over the initial phase. 
Equations (2) are effectively one-dimensional, but $\vec{q}$ and $\W$ are 
two-dimensional (2D) parameters and 
this 2D dependence will prove to be of vital importance.

In order to enhance the bandwidth of the FEL gain, we consider
the setup of two identical wigglers of equal length $L_W$ with a specially
designed drift region between them, as described below.
Since the change of electron energy inside the first and second wigglers is
given by the same set of equations (2)-(4), we obtain,
by taking into account the
phase shift 
%(\ref{ph-shift}) 
in the drift region and  averaging 
over the random initial phases,
the following expression for gain in the entire two-wiggler setup 
\cite{nikonov96pre,nikonov98pre} 
\begin{eqnarray}
\mbox{Gain}
\sim<\Delta\gamma>
\sim -<\Delta \Omega >&=&{\frac 1{\Omega ^3}}\Bigl[2\Omega T\sin \Omega
T+4\cos \Omega T-4+ \nonumber \\ & &
+2\Omega T\sin (2\Omega T+\Delta \psi )-2\Omega T\sin (\Omega T+\Delta \psi )
\nonumber \\ & &
+2\cos \Delta \psi +2\cos (2\Omega T+\Delta \psi )-4\cos (\Omega T+\Delta
\psi )\Bigr].
\end{eqnarray}
where $T=L_W/c(1-v_z/c)$ is the mean interaction time in the wiggler.

For $\Delta\psi=0$ the two-wiggler gain coincides with the
result for one wiggler of twice the length. The resulting gain dependence
on the detuning $\W$ 
and laser frequency variation $\delta \omega$
is depicted in Fig.~\ref{fel-dw}a, 2b, respectively.
The average of this antisymmetric gain 
over detuning vanishes, 
in accordance with the Madey gain-spread theorem, 
which is the main restriction on gain in short-wavelength FELs \cite{FEL}.

The electron oscillates coherently in the ponderomotive potential, and
therefore its oscillations in the two sequential wigglers exhibit
interference with a phase $\Delta\psi$ 
which depends on the path (or time) difference between the two
regions. 
In an optical klystron, the phase shift, 
produced in a free space of length $L$ between
the wigglers, is equal to 
\begin{equation}
\Delta \psi_{\rm klystron} =k_WL+{\frac{ck_Lq_zL}{\omega ^2}}\Omega .
\end{equation}
The gain dependence of the optical klystron on the detuning $\Omega$ and laser
frequency $\delta \omega$ is
depicted in Fig.~\ref{phok}a. The maximum gain 
exceeds that of 
the ordinary FEL, but the restriction 
on the spread of $\W$ (and therefore on $\delta\omega$ in (5))
becomes more stringent, because of rapid oscillations of the
gain dependence on the detuning (Fig.~\ref{phok}a).

\subsection{Broadband gain by drift-region optical dispersion}

In order to overcome the limitations 
of FELWI phase-shift implementation for electrons in magnetic fields 
discussed in Sec. I, 
we consider the alternative phase shift produced by an {\it optically 
dispersive} drift region, where the light path depends 
on the deviation $\delta\omega$ of the laser frequency as follows
\be
\Delta\psi = \Delta\psi_0 + \delta\omega 
\ds  \left(
{s_L(\omega) + \omega\ds{d s_L(\omega)\over d\omega}\over c}-{s_e\over v} 
\right).
\label{opt-disp}
\ee
Here $s_L(\omega)$ is the optical path depending on 
the laser frequency in the drift region, $s_e$ 
is the electron path passing through the drift 
region as before, $\omega_0$ is the mean laser frequency, 
and the corresponding phase shift is equal to
\be
\Delta\psi_0 = \omega_0({s_L(\omega_0)\over c} - {s_e \over v}).
\ee

The spectral dispersion of 
the drift region in (\ref{opt-disp}) 
allows us to manipulate the dependence of the gain (6) on detuning. 
As a result, {\it broadband gain} appears, as shown 
in Fig.~3b. 
The explicit condition which 
the dispersion (\ref{opt-disp}) must satisfy in order 
to obtain broadband gain is
\be
{s_L(\omega) + \omega\ds{d s_L(\omega)\over d\omega}\over c} 
\simeq {s_e\over v} .
\label{e-14}
\ee

The gain dependence on the laser frequency deviation $\delta \omega$ exhibits a
broad band in Fig.~3b 
 for the same  parameters as in Fig.~3a: 
the contrast between our design and
 an optical klystron is striking indeed, and demonstrates the crucial 
effect of optical dispersion on gain. 

The extension of this treatment to situations wherein the effects of
$\delta v_z$ and $\delta \omega$ are comparable involves the averaging of (6)
and (8) over a thermal spread of electron velocities. For moderate spreads
the broadband character of the gain persists, as seen in Fig. \ref{phok}b.

Before we discuss implementations of the drift region 
for broadband gain, 
let us note here that, although, the current and the FEL WI 
concepts are both based upon interference of radiation emitted 
by electron moving in the first and the second wigglers via phase shift 
created by the drift region, there is an important difference. 
As has been shown in \cite{nikonov98pre}, the motion in the 
FEL WI drift region has to be 2D (otherwise, the phase density is conserved 
in accordance with the Liouville's theorem, and, for electron distribution 
having large spread of momenta, the gain is zero).
To obtain the broadband gain, 
it is not necessary to have the setup of drift region
2D (for electron motion),
but rather the optical dispersion of the drift region
should satisfy to Eq.~(\ref{e-14}).

\section{Drift region designs for broad band gain}

We shall  now discuss two  possible experimental implementations
of optically dispersive drift regions created by optical 
elements in order to get broadband gain: (a) diffraction by prisms; 
(b) Bragg reflectors.

\subsection{Drift region dispersion  based on prisms}

Let us consider a setup of the drift region depicted in Fig.~\ref{dr-1}.  
After the first wiggler the electron beam, having passed through 
free space of 
length $s_e$, enters the second wiggler. The laser beam is diffracted 
and guided by the set of prisms 1, 2, 3, and 4, which are adjusted 
to have {\em vanishing total dispersion}. 
The phase shift introduced by this setup for the laser field is given by
\begin{equation}
\Delta \psi =k_L\left( 2x_0\left( {\frac 1{\cos \alpha }}-1\right)
-s_e\left( {\frac c{v_e}}-1\right) \right),
\end{equation}
where $x_0$ is the distance between prisms 1 and 2, 
$\alpha $ is the angle of diffraction for the laser field,
$\pi /2$ is the tip angle of the prism,  
$s_e$ is the distance between prisms 1 and 4, which 
is also the electron path in the drift region. For a
small deviation of the laser frequency $\delta \omega$, 
the phase shift is given by 
\begin{equation}
\Delta \psi =\Delta \psi _0+\displaystyle {\frac 1c}\left(s_e\left( 1-{%
\frac c{v_e}}\right) +2x_0\left( {\frac 1{\cos \alpha }}-1\right) +{\frac{%
2x_0\omega \displaystyle{\frac{d\alpha }{d\omega }}\tan \alpha }{\cos \alpha 
}}\right) \delta \omega ,
\label{e-16}
\end{equation}
where the phase shift for the central frequency $\omega_0$ is given by 
\begin{equation}
\Delta \psi _0=\displaystyle {\frac{\omega_0}{c}}\left( s_e\left( 1-{\frac c{v_e}%
}\right) +2x_0\left( {\frac 1{\cos \alpha }}-1\right) \right),
\end{equation}
$\ds{d\alpha\over d\omega}$ being the angular dispersion of the prism. 

A properly chosen dispersion of the prisms,
in accordance with Eq.~(\ref{e-14}),
 allows us 
to adjust the drift region so as to have 
the broadband optical gain (see Fig.~3b). The choice,
which cancels the $\delta \omega$ term in (\ref{e-16}), is: 
\be 
\frac{d\alpha }{d\omega } = \ds{
s_e\left( {\frac c{v_e}} - 1\right) + 
2x_0\left( 1 - {\frac 1{\cos \alpha }}\right) 
\over 2x_0\omega \tan \alpha} \cos \alpha .
\ee

\subsection{Drift region dispersion based on a Bragg reflector}

The system of prisms is not the only way to create 
a proper phase shift. We may use a Bragg reflector instead,  
depicted in Fig.~\ref{dr-2}.
The phase shift introduced by a Bragg reflector is given by
\be
\psi = \tan^{-1}\left(
{1 - R \over 1 + R}{1\over \tan\ds\left({kl\over 2\cos\theta}\right) }
\right),
\ee
where $R$ is the reflectance 
of the Bragg structure, $l$ is its length, $\theta$ is the angle of incidence.
%The derivative of the phase shift is given by
The broadband gain conditions for this setup are cos$\theta > \frac{v}{c}$ and
\begin{eqnarray} 
\frac{d\psi}{d\omega} = 
{1 - R^2 \over 
\ds\sin^2\left({kl\over 2\cos\theta}\right) (1+ R)^2 +
\ds\cos^2\left({kl\over 2\cos\theta}\right) (1- R)^2}
\ds{l\over 2 c \cos\theta} 
=
L_e \left(
\ds{1\over v} - \ds{1\over c\cos\theta}  
\right),
\end{eqnarray}
where $L_e$ is the distance between points mirrors 1 and 2 (Fig.~\ref{dr-2}).

It is instructive to calculate the dependence of gain width on the parameters
of the drift regions.  
In Fig.~\ref{width}, 
we show the gain-width dependence on the dispersion of the drift region,
$d\psi/d\omega$. 
The figure demonstrates that the gain width can be made as large as the 
spectral range wherein the pondermotive interaction between the electrons and 
the laser field does not vanish
(as seen in Fig.~3a, 
there is no spectral range where the gain is negative).
Clearly, the parameters that provide broad-band gain are different 
for the drift regions shown in Fig.~\ref{dr-1} and Fig.~\ref{dr-2},
but the gain-width dependence on the total dispersion $d\psi/d\omega$ 
is {\em universal\/}.

\section{Conclusions}

We have shown in this paper that the small-signal gain 
in a FEL comprised of two coupled wigglers can exhibit 
a broadband character as a function of the laser frequency, if their 
interface (the region wherein electrons and light drift 
without interaction) is endowed with appropriate optical dispersion. 
This design  is based on the same principles 
as in \cite{sherman95prl,nikonov96pre}, 
except that in \cite{nikonov96pre} specially designed 
magnetic fields are proposed for manipulating 
the drift-region phase shifts, whereas here 
linear optical elements, such as prisms (Sec.~IIIA) 
or a Bragg reflector (Sec.~IIIB) suffice. 
The present dispersive scheme, similarly to \cite{nikonov96pre}, 
achieves the cancellation of absorption in a broad 
spectral range by destructive interference of frequencies 
that contribute on average to loss in the two wigglers, 
and the reinforcement of emission by constructive interference 
of frequencies that contribute to gain. 
Remarkably, the resulting broad-band gain exhibits {\em universal\/} 
dependence on the optical dispersion, and is compatible with existing FELs,
regardless of their design. The broadband character of the gain persists for
moderate spreads of electron velocities.

The proposed scheme may allow 
effective FEL operation using femtosecond 
optical pulses, which correspond to a wide
spectral band: a gain bandwidth of $\sim 20c/L_W$ (Fig. 3) may attain $10^{14}$
Hz values for optical wigglers with submicron periods 
\cite{FEL,Nikon,Madeythe}.

\acknowledgments

The authors  gratefully acknowledge 
support from the Office of Naval Research,
the National Science Foundation, the Robert A. Welch Foundation, 
and the US-Israel BSF.

\newpage

\section*{Figures}

%\newpage
\begin{figure}%[tbp]
%\center{\epsfig{file=fig1bok.eps,  width=15cm, angle=0} }
\label{ifel}
\caption{Two coupled wigglers separated by a drift region.}
\end{figure}

%\vspace*{2cm} Fig.~1, 
%Y. Rostovtsev, et al., ``Broadband optical gain via ...''

%\newpage
\begin{figure}[tbp]
%\center{
%\epsfig{file=p-fel/f-fel-dw.eps, width=10cm, angle=0} }
\caption{Gain (arbitrary units) dependence on detuning $\Omega$ as a
function of (a) $\delta v_z = v_z - \omega / (k_W+k_L)$ and (b)
laser frequency variation $\delta \omega$, for
the ordinary free electron laser (FEL). The frequency variation is normalized
by the interaction time in the wigger $T=L_W/c(1-v_z/c)$.}
\label{fel-dw}
\end{figure}

%\vspace*{2cm} Fig.~2, 
%Y. Rostovtsev, et al., ``Broadband optical gain via ...''

%\newpage
\begin{figure}[tbp]
%\center{\epsfig{file=p-fel/f-ok-dw.eps, width=10cm, angle=0} }
\caption{
Gain (arbitrary units) dependence on laser frequency for 
two coupled wigglers. 
(a) As is clearly seen, in comparison with Fig.~2, 
the maximum gain for an optical klystron configuration 
has been increased by adding a proper phase shift via 
drift region but without adjustment of dispersion for laser beam. 
The gain dependence on the laser frequency experiences fast oscillations
because of interference between waves emitted 
in the first and in the second wiggler.  
(b) By proper adjustment of the phase shift for every laser frequency 
via an appropriately designed dispersive drift region, 
satisfying Eqs. (8)-(10), the broad band gain has been obtained. 
For the case (b) two curves are shown: (1) without electron momentum spread and
(2) with electron velocity spread $\Delta v_z = 5c/L_W(k_L+k_W)$. 
Note the broad gain bandwidth, which allows for 
ultrashort pulsed FEL operation ($\delta\omega \sim 20 c/L_W$)
(normalization by $T$ as in Fig. \ref{fel-dw}).
}
\label{phok}
\end{figure}

%\vspace*{2cm} Fig.~3, 
%Y. Rostovtsev, et al., ``Broadband optical gain via ...''

%\newpage
\begin{figure}[tbp]
%\center{\epsfig{file=broadband-ok.eps,  width=12cm, angle=0} }
\caption{Scheme of the drift region with prisms for a broadband FEL (or 
optical phased klystron).}
\label{dr-1}
\end{figure}

%\vspace*{2cm} Fig.~4, 
%Y. Rostovtsev, et al., ``Broadband optical gain via ...''

%\newpage
\begin{figure}[tbp]
%\center{\epsfig{file=bragg-ok.eps,  width=12cm, angle=0} }
\caption{Scheme of the drift region with a Bragg reflector for a broadband FEL
.}
\label{dr-2}
\end{figure}

%\vspace*{2cm} Fig.~5, 
%Y. Rostovtsev, et al., ``Broadband optical gain via ...''

%\newpage
\begin{figure}[tbp]
%\center{\epsfig{file=gainwidth.eps, 
%width=10cm, angle=0} }
\caption{Dependence of the FEL gain width on the dispersion parameter 
$d\psi/d\omega$. 
It is seen that gain exists for all FEL frequencies at which 
strong interaction with the pondermotive potential occurs.
The same type of dependence appears for drift regions shown in Figs.~4,5. 
The parameter space is different, 
but in both cases there is a parameter region where broad-band gain exists
(normalization by $T$ as in Fig. \ref{fel-dw}, \ref{dr-1}). 
}
\label{width}
\end{figure}

%\vspace*{2cm} Fig.~6, 
%Y. Rostovtsev, et al., ``Broadband optical gain via ...''

\end{document}